\documentclass[preprint,12pt,3p]{elsarticle}
\usepackage{amssymb}
\usepackage{booktabs}
\usepackage{dcolumn}
\usepackage{multirow}
\usepackage{adjustbox}
\usepackage{subfig}
\usepackage{graphicx}
\usepackage{natbib} 
\usepackage{booktabs}
\usepackage{lscape}
\usepackage{amsmath}
\usepackage{float}
\usepackage[obeyspaces,hyphens]{url}
\usepackage{xcolor}
\usepackage{empheq}
\usepackage[countmax]{subfloat}
\biboptions{sort&compress}
\newcommand{\be}{\begin{equation}}
\newcommand{\ee}{\end{equation}}
\newcommand{\beq}{\begin{eqnarray}}
\newcommand{\eeq}{\end{eqnarray}}
\newcommand{\lb}[1]{\label{#1}}

\newcommand{\ssty}{\scriptscriptstyle}

\newcommand{\dd}{{\mathrm{d}}}
\newcommand{\dl}{d_{\ssty L}}

\newcommand{\gaml}{\gamma_{\ssty L}}

\newcommand{\barbeta}{{\bar{\beta}}}
\makeatletter
\newcommand{\oz}{\mathbin{\mathpalette\make@circled{z}}}
\newcommand{\odl}{\mathbin{\mathpalette\make@circled{\dl}}}
\newcommand{\ogams}{\mathbin{\mathpalette\make@circled{\gaml^\ast}}}
\newcommand{\make@circled}[2]{%
  \ooalign{$\m@th#1\smallbigcirc{#1}$\cr\hidewidth$\m@th#1#2$\hidewidth\cr}%
}
\newcommand{\smallbigcirc}[1]{%
  \vcenter{\hbox{\scalebox{1.6}{$\m@th#1\bigcirc$}}}%
}
\makeatother
\usepackage{array}

\journal{}

\begin{document}


\begin{frontmatter}

\title{Matching the Alcubierre and Minkowski spacetimes}

\author[1]{Osvaldo L.\ Santos-Pereira\fnref{fn1}}
\address[1]{Physics Institute, Universidade Federal do Rio de Janeiro, Rio
	de Janeiro, Brazil}
\ead{olsp@if.ufrj.br}

\author[2,3]{Everton M.\ C.\ Abreu\fnref{fn2}}
\address[2]{Physics Department, Universidade Federal Rural do Rio de Janeiro,
        Serop\'{e}dica, Brazil}
\address[3]{Graduate Program in Applied Physics, Universidade Federal do Rio
        de Janeiro, Brazil}
\ead{evertonabreu@ufrrj.br}

\author[1,3]{Marcelo B.\ Ribeiro\fnref{fn3}\corref{cor3}%
}
\ead{mbr@if.ufrj.br}
\cortext[cor3]{\it Corresponding author}
\fntext[fn1]{Orcid 0000-0003-2231-517X}
\fntext[fn2]{Orcid 0000-0002-6638-2588}
\fntext[fn3]{Orcid 0000-0002-6919-2624}

\begin{abstract}
This work analyzes the Darmois junction conditions matching an
interior Alcubierre warp drive spacetime to an exterior Minkowski
geometry. The joining hypersurface requires that the shift vector of
the warp drive spacetime must satisfy the solution of a particular
inviscid Burgers equation, namely, the gauge where the shift vector is
not a function of the $y$ and $z$ spacetime coordinates. Such a gauge
connects the warp drive metric to shock waves via a Burgers-type
equation, which was previously found to be an Einstein equations vacuum
solution for the warp drive geometry. It is also shown that not all
Ricci and Riemann tensors components are zero at the joining
hypersurface, but for that to happen they depend on the shift vector
solution of the inviscid Burgers equation at the joining wall. This
means that the warp drive geometry is not globally flat.
\end{abstract}


\end{frontmatter}


\section{Introduction}

Alcubierre \cite{Alcubierre1994} proposed a propulsion mechanism based
on General Relativity capable of transporting massive particles at
superluminal speeds by positioning them inside a spacetime distortion
formed by a special asymptotically flat metric. The function that in
this geometry describes the mass particle transportation distortion was
called the \textit{shift vector}. In reference to science fiction
literature he named this propulsion system \textit{warp drive} (WD),
then the distorted, or warped, spacetime defined by the shift vector can
be likewise called the \textit{warp bubble}, and the global superluminal
velocities are similarly called \textit{warp speeds}.

Alcubierre's work \cite{Alcubierre1994} sparked general interest that
led to what is now an extensive literature, so far mostly discussing
the violation of weak, null, strong, and dominant energy conditions
\cite{Olum1998,Visser1999,Lobo2002,LoboVisser2004,Lobo2007,
AlcubierreLobo2017,Santiago2021a,Santiago2021b}. Other works proposed
the use of quantum energy inequalities and quantum effects from
semi-classical and quantum field theories to discuss the possibility
of creating a WD framework \cite{Hiscock1997,Pfenning1997,FordRoman1996,
Vollick1997,Pfenning1998,Roman1998,Krasnikov2003,Ellis2004,Quarra2016}.
Some authors attempted to circumvent the requirement of negative energy
for creating superluminal speeds by proposing different WD spacetime
metrics \cite{Krasnikov1998,EveretRoman1997,Natario2002,Fischer2002,
Palmer2003,LoboVisser2004,Gravel2004,Varieschi2013,Lobo2016}, which
included the construction of gravitational tubes \cite{Krasnikov1998,
Krasnikov2003}, conformal expansion terms, a linearized WD metric with
a type of Schwarzschild potential, and even a spacetime with no bubble
volume expansion. Related works followed these new proposed geometries,
recalculating the negative energy density required to create the WD
superluminal speeds effect \cite{Broeck1999, Loup2001}. 

Refs.\ \cite{nos1,nos2,nos3,nos4,nos5} followed a somewhat different
approach by proposing to couple the WD geometry to simple known
sources of matter and energy, and then solve the Einstein equations
to see what matter-energy requirements and constraints appear on the
WD functions, particularly on the shift vector, once the Einstein
equations are imposed upon the WD metric. Ref.\,\cite{nos1} solved
the Einstein equations for the WD metric having dust as source in
the energy-momentum tensor, and the respective solutions turned out
to be a vacuum. Moreover, the solutions connected the WD geometry to
shock waves via a Burgers-type equation that also came out of the
solutions, which suggested that the warp bubble might be a shock wave
moving in a flat spacetime, further indicating a physical limit between
the warp bubble and the flat Minkowski background.

One possible way of describing the warp bubble as embedded in a flat
spacetime is to join the WD metric to the Minkowski background by means
of \textit{junction conditions} in an attempt to reveal the physical
constraints imposed upon the interior geometry by such a match. That
would be similar to what has already been previously studied is
cosmology regarding the limitations of the Lema\^{\i}tre-Tolman-Bondi
cosmology overdensity and underdensity once it is inserted as a
spherical region inside a Friedmann-Lema\^{\i}tre-Robertson-Walker
standard cosmological model background \citetext{\citealp{Ribeiro1992},
$\mathsection$4; \citealp{krasinski1997}, pp.\ 136; \citealp{krasinski2006},
$\mathsection$18.13}, or the possible influence of the expansion of the
Universe in the Solar System using the Einstein-Straus configuration,
formed by a Schwarzschild vacuum embedded in a Friedmann or
Minkowski background \citetext{\citealp{krasinski1997}, $\mathsection$3.3;
\citealp{krasinski2006}, $\mathsection$18.7}. In the WD context, an
approach using junction conditions to examine the limits of the warp
bubble was studied in Ref.\ \cite{venezuelanos}, although these authors
analyzed the problem using a static, not warped, bubble, an approach
somewhat different from \textit{Warp Drive Theory} in the sense of 
Alcubierre because a static bubble does not entail warp speeds.

In this work the junction conditions between the WD and Minkowski metrics
are investigated. Is is shown that the gluing of these two metrics are
continuous under the gauge
\be
(\partial \beta/\partial y)^2+(\partial \beta/\partial z)^2=0,
\lb{gauge}
\ee
where $\beta$ is the WD shift function. This gauge also leads to the
energy density being equal to zero, which is the case where the
Burgers-type equation satisfies the vacuum Einstein equations
\cite{nos1}. The shift vector $\beta$ carries the basic information
on how the WD spacetime behaves and the kinematics of the observer
with Eulerian world lines moving in space. It is also shown that
the Riemann and Ricci tensors are not identically zero, but depend
on the solution of the inviscid Burgers equation, demonstrating
that the WD spacetime is not globally flat.

The plan of this paper is as follows. Sec.\,\ref{wdvsminsk} calculates
the Darmois junctions conditions between WD and Minkowski spacetimes,
demonstrating that both of them can be continuously joined when
considering the results in Refs.\ \cite{nos1, nos2, nos4, nos5}. We
discuss the results considering two shift vectors, the original one
proposed by Alcubierre $\beta$, and the one used in the just cited works,
defined as $\bar{\beta} = -\beta$. Sec.\ 3 shows that the Riemann and
Ricci tensors are only zero when the inviscid Burgers equation is
satisfied for $\beta$ and $\bar{\beta}$. In addition, the original
WD $\beta$ the conditions for flatness are points where a linear
combination solution of the heat equation and the viscous Burger equation
are satisfied.  Sec.\ \ref{conc} presents our conclusions.

\section{Matching the Warp Drive and Minkowski spacetimes}\lb{wdvsminsk}

Let $V^-$ and $V^+$ be two spacetime regions separated by the
hypersurface $\Sigma$, where $V^-$ refers to the interior region and
$V^+$ to the exterior one. Then $x^\mu_{-}$ and $x^\nu_{+}$ are the
coordinates of each respective region, and $g_{\mu\nu}^-$ and
$g_{\mu\nu}^+$ the corresponding metrics. Greek indices $(\mu=0,1,2,3)$
refer to 4-dimensional regions, whereas Latin indices $(a=0,2,3)$ refer
to the joining 3-dimensional hypersurface $\Sigma$ whose specific
coordinates will be defined below.

Let us now consider the exterior spacetime $V^+$ as being the
Minkowski metric, which may be written as below,
\be
\dd s^2 = - \dd T^2 + \dd X^2 + \dd Y^2 + \dd Z^2.
\lb{mink}
\ee
The interior region $V^-$ will then be the WD spacetime, but its
shift vector may be written with different signs, as follows,
\be
\bar{\beta} = - \beta. 
\label{newshift}
\ee
The shift vector is fundamental in warp drive dynamics because it is the
actual generator of warp speeds. Alcubierre \cite{Alcubierre1994}
originally advanced $\beta$, whereas $\bar{\beta}$ was studied in Refs.\
\cite{nos1,nos2,nos3,nos4,nos5}. Each sign of the shift vector has a
different physical significance and leads to different dynamics
\cite[see][for details]{nos6}, so for this reason here we shall study
both cases, starting with~$\bar{\beta}$.

\subsection{Shift vector $\bar{\beta}$}

It follows from Refs.\ \cite{nos1,nos2,nos3,nos4,nos5} that the
interior WD metric may be written as below,
\be
\dd s^2=-\left(1-{\bar{\beta}}^2\right)\dd t^2 - 2 {\bar{\beta}}
\, \dd x \, \dd t + \dd x^2 + \dd y^2 + \dd z^2.
\lb{ozwdmetric}
\ee
The joining hypersurface $\Sigma$ may be defined on each side of the two
spacetimes as follows, 
\begin{subequations}
\begin{empheq}[left=\empheqlbrace]{align}
&\Sigma_-(x^a_-) = x-\Sigma_0= 0, \lb{regionin} \\[-10pt] \nonumber \\ 
&\Sigma_+(x^a_+) = X-\Sigma_0= 0,  \lb{regionout} 
\end{empheq}
\end{subequations}
where $\Sigma_0$ is a constant. The junction metric $\dd s^2_\Sigma$ on
$\Sigma$ yields, 
\be
\dd s^2_{\Sigma} = g_{ab} \, \dd\xi^a \dd\xi^b,
\label{firstjc}
\ee
where $\xi^a$ are the intrinsic coordinates on $\Sigma$. For the match
to happen between the geometries given by Eqs.\ \eqref{mink} and
\eqref{ozwdmetric} the \textit{first fundamental form} $\dd s^2$ of
these two spacetimes must be identical on $\Sigma$, that is, $\dd s_-^2=
\dd s_+^2$, which means $\xi^a=x_-^a= x_+^a$. Then the expressions below
are straightforward,
\begin{subequations}
\begin{empheq}[left=\empheqlbrace]{align}
&\xi^0=T=t\sqrt{1-{\bar{\beta}}^2}, \lb{ksi0} \\[-15pt] \nonumber \\
&\xi^2=Y=y, \lb{ksi2} \\[-15pt] \nonumber \\
&\xi^3=Z=z. \lb{ksi3}
\end{empheq}
\end{subequations}
Eq.\ \eqref{ksi0} implies that
$1=\sqrt{{\xmathstrut{0.05}}{1-{\barbeta}^2}}$
on $\Sigma$, which allows us to reach at the \textit{first junction
condition}, 
\be
\bar{\beta}=0, \;\;\; \mathrm{on} \;\;\; \Sigma.
\lb{jc1}
\ee

This result is not surprising, since this is the condition on which
the WD metric \eqref{ozwdmetric} turns into Minkowski. However, it 
should be emphasized that this is mandatory only over the joining
hypersurface $\Sigma$, which means that $\bar{\beta}$ can be different
from zero inside the interior region, or obey other conditions for the
spacetime match to take place. In other words, this is not the solely
junction condition which means that the shift vector may be such that
the match could take place without a vanishing $\bar{\beta}$, as we
shall see below.

The junction conditions also require that the \textit{second fundamental
form}, the extrinsic curvature $K_{\mu\nu}$, must match on $\Sigma$,
which means that the condition $K^-_{ab}=K^+_{ab}$ must be satisfied. One
calculates this according to the definition of $K_{\mu\nu}$ on both
regions in their respective coordinates $x_\pm^\mu$ and then having a
coordinate transformation to the joining hypersurface coordinates
$\xi^a$. Hence, the extrinsic curvature
\be
K_{\mu\nu}=-n_{\mu;\nu}
\lb{kmunu}
\ee
takes on $\Sigma$ the form below,
\be
K_{ab}=\frac{\partial x^\mu}{\partial \xi^a} \frac{\partial x^\nu}
{\partial \xi^b} K_{\mu\nu},
\lb{extrin1}
\ee
where $n_{\mu}$ are the normal unit vectors over $\Sigma$ pointing
inward (minus label) and outward (plus label) as follows,
\be
n_\mu^\pm = \frac{\Sigma^\pm,_{\mu}}
{\sqrt{g^{\alpha\sigma}\Sigma^\pm,_{\alpha}\Sigma^\pm,_{\sigma}}}. 
\lb{normalunit}
\ee
The normal vector on the Minkowski side is constant, which means the 
trivial and straightforward result that all components of $K^+_{ab}$ on
$\Sigma$ vanish. This implies that the junction conditions required
for the second fundamental form are reduced to calculating the nonzero
components of $K^-_{ab}$ and equating them to zero, that is, 
\be
K^-_{ab}=0, \;\;\; \mathrm{on} \;\;\; \Sigma.
\lb{jc2}
\ee
The extrinsic curvature tensor \eqref{extrin1} may be rewritten as
follows,
\be
K_{ab} = - n_{\mu;\nu} \, e^\mu_a \, e^\nu_b= (n_{\mu,\nu}
+ \Gamma^\alpha_{\mu\nu} n_\alpha)  e^\mu_a \, e^\nu_b,
\lb{extrin2}
\ee
where
\be
e^\mu_a = \frac{\partial x^\mu}{\partial \xi^a}.
\lb{tangent}
\ee
are tangent vectors calculated on points of the hypersurface whose
projections to the normal vector onto $\Sigma$ are always zero,
\be
e^\mu_a\,e^\nu_b\,n_{\mu},_\nu = 0.
\lb{proj1}
\ee
Then Eq.\ \eqref{extrin2} is reduced to,
\be
K_{ab} = e^\mu_a\,e^\nu_b\,\Gamma^\sigma{}_{\mu\nu}\, n_\sigma .
\label{extr1}
\ee
Now, remembering Eq.\ \eqref{normalunit} it follows that the joining
hypersurfaces defined by Eqs.\ \eqref{regionin} and \eqref{regionout}
yield,
\be
n_\mu = (0,1,0,0).
\label{normal1}
\ee
and then the only nonzero terms for Eq.\ \eqref{extr1} are the ones
having $\Gamma^1{}_{\mu\nu}$. Since ${(\Gamma^1{}_{\mu\nu})}^+=0$, the
only connection components on $V^-$ relevant for calculating the
extrinsic curvature are the ones below, 
\begin{align}
{(\Gamma^1{}_{00})}^{-} &= -\frac{\partial \barbeta}{\partial t}
+(\barbeta^3 -\barbeta)
\frac{\partial \barbeta}{\partial x}, \lb{gama100} \\
{(\Gamma^1{}_{02})}^{-} &= -\frac{1}{2}(1+{\barbeta^2})\frac{\partial \barbeta}
{\partial y}, \lb{gama102} \\
{(\Gamma^1{}_{03})}^{-} &= -\frac{1}{2}(1+{\barbeta^2})\frac{\partial \barbeta}
{\partial z}. \lb{gama103}
\end{align}
Then, according to Eq.\ \eqref{jc2} the nonzero extrinsic curvature
components on $V^-$ yield,
\begin{align}
K_{00}^- &= \frac{\partial\bar{\beta}}{\partial t} + \left(\bar{\beta}
            - \bar{\beta}^3 \right) \frac{\partial \bar{\beta}}{\partial x}
	    = 0, \label{juncond1} \\
K_{02}^- &= \frac{1}{2}(1 + \bar{\beta}^2) \frac{\partial\bar{\beta}}
	    {\partial y} = 0, \label{juncond2} \\
K_{03}^- &= \frac{1}{2}(1 + \bar{\beta}^2) \frac{\partial\bar{\beta}}
	    {\partial z} = 0. \label{juncond3}
\end{align}
Eqs.\ \eqref{juncond1} to \eqref{juncond3} form the
\textit{second group of junction conditions} gluing the WD metric to
Minkowski's one. They may be written as below,
\begin{subequations}
\begin{empheq}[left=\empheqlbrace,right=\empheqrbrace\;\mathrm{on}\;\;\Sigma.]
	{align}
&\frac{\partial\bar{\beta}}{\partial t}+\left(\bar{\beta}-\bar{\beta}^3 \right)
\frac{\partial \bar{\beta}}{\partial x}=0, \lb{result1} \\[-8pt]
\nonumber \\
&\frac{\partial\bar{\beta}}{\partial y}=\frac{\partial\bar{\beta}}{\partial z}
=0, \lb{result2} \\
\nonumber \\
&\bar{\beta} = \pm i, \label{result3}
\end{empheq}
\end{subequations}
Although a purely imaginary $\barbeta$ may in principle be regarded as
unphysical, it is included here for completeness. One may recall that the
first junction condition is given by Eq.\ \eqref{jc1} above. 

In order to analyze the results above, let us first remember the general
form of the Burgers equation \cite[$\mathsection\mathsection$3.4, 4.4]
{evans},
\be
\frac{\partial u}{\partial t}+c(u)\frac{\partial u}{\partial x}= \nu
\frac{\partial^2 u}{\partial x^2},
\lb{generalburgers}
\ee
where $c(u)$ is a general function of the velocity vector field $u(t,x)$
and $\nu$ is the diffusion term. So Eq.\ \eqref{result1} may be interpreted
as a general inviscid Burgers equation \cite{Burgers1948, Cole1951} having
$\nu=0$ and whose general function $c(\bar{\beta})$ is given by,
\be
c(\bar{\beta}) = \bar{\beta} - \bar{\beta}^3.
\lb{cbeta}
\ee
Eq.\ \eqref{generalburgers} is a quasilinear hyperbolic equation if the
condition $c(\bar{\beta}) > 0$ is satisfied, and its solution can be
constructed using the method of characteristics. This condition implies
that for this specific regime of the Burgers equation the shift vector
must obey the following inequality,
\be
0 < |\bar{\beta}| < 1, \;\;\; \mathrm{on} \;\;\; \Sigma.
\lb{inequa}
\ee

Hyperbolicity in partial differential equations is often associated to
wave like behavior and possesses real characteristics, propagating information 
along the characteristic curves, which means that a particle inside the 
warp bubble may present wave like behavior. Notice, however, that the
condition \eqref{inequa} is for the possibility of wave like behavior. It
does not mean that the solution must behave in this way.

Let us now assume that the inviscid Burgers equation having
$c(\bar{\beta})=\bar{\beta}$ is satisfied on $\Sigma$. Then Eq.\
\eqref{result1} can be expressed as below,
\be
\frac{\partial\bar{\beta}}{\partial t} + \bar{\beta} \frac{\partial
\bar{\beta}} {\partial x} = \bar{\beta}^3 \frac{\partial \bar{\beta}}
{\partial x} = 0, 
\lb{result1-a}
\ee
from where it is straightforward to see that either $\bar{\beta} = 0$ or
$\partial \bar{\beta}/\partial x = 0$. The former is consistent with the
junction condition \eqref{jc1}, whereas the latter
implies that $\partial \bar{\beta}/\partial t = 0$ and then $\bar{\beta}$
being constant.

Finally, one may notice that Eqs.\ \eqref{result2} are equivalent to
the gauge in Eq.\ \eqref{gauge} for the shift vector $\bar{\beta}$
found in the vacuum solutions of the Einstein equations connecting the
WD to shock waves via Burgers-type equations \cite[see Refs.][]{nos1,
nos2,nos3,nos4,nos5}.

\subsection{Alcubierre shift vector $\beta$}\lb{menosbeta}

The original Alcubierre WD metric \cite{Alcubierre1994} may be written
as below, 
\be
\dd s^2=-\left(1-{\beta}^2\right)\dd t^2 + 2 {{\beta}}
\, \dd x \, \dd t + \dd x^2 + \dd y^2 + \dd z^2.
\lb{alcmetric}
\ee
Following the same prescription used in the previous section to
calculate the junction conditions for the interior WD and the exterior 
Minkowski spacetime we arrived at similar results. The first fundamental
form results in,
\be
{\beta}=0, \;\;\; \mathrm{on} \;\;\; \Sigma.
\lb{jc3}
\ee
The normal vector is the same as in Eq.\ \eqref{normal1}. The relevant
connection components read,
\begin{align}
{(\Gamma^1{}_{00})}^{-} &= \frac{\partial \beta}{\partial t}
+(\beta^3-\beta) \frac{\partial \beta}{\partial x}, \lb{alcgama100} \\
{(\Gamma^1{}_{02})}^{-} &= \frac{1}{2}(1+{\beta^2})\frac{\partial \beta}
{\partial y}, \lb{alcgama102} \\
{(\Gamma^1{}_{03})}^{-} &= \frac{1}{2}(1+{\beta^2})\frac{\partial \beta}
{\partial z}, \lb{alcgama103}
\end{align}
and the extrinsic curvature components yield, 
\begin{align}
K_{00}^- &= \frac{\partial\beta}{\partial t} + \left(\beta^3 - \beta
\right) \frac{\partial \beta}{\partial x} = 0, \label{wjuncond1} \\
K_{02}^- &= \frac{1}{2}(1 + \beta^2) \frac{\partial\beta}{\partial y} = 0,
\label{wjuncond2} \\
K_{03}^- &= \frac{1}{2}(1 + \beta^2) \frac{\partial\beta}{\partial z} = 0.
\label{wjuncond3}
\end{align}
Hence, the second group of junction conditions becomes,
\begin{subequations}
\begin{empheq}[left=\empheqlbrace,right=\empheqrbrace\;\mathrm{on}\;\;\Sigma.]
	{align}
&\frac{\partial\beta}{\partial t} + \left(\beta^3 - \beta \right)
\frac{\partial \beta}{\partial x} = 0, \label{wresult1} \\[-8pt]
\nonumber \\
&\frac{\partial\beta}{\partial y} = \frac{\partial\beta}{\partial z}
= 0, \label{wresult2} \\
\nonumber \\
&\beta = \pm i, \label{wresult3}
\end{empheq}
\end{subequations}
which are similar to the ones found in the previous section, apart from
the signal symmetry in Eqs.\ \eqref{result1} and \eqref{wresult1}. The
same gauge $\partial\beta/\partial y=\partial\beta/\partial z = 0$ for
the Einstein-Burgers vacuum solution \cite{nos1,nos2,nos3,nos4,nos5}
also appears, and again the purely imaginary result $\beta = \pm i$ is
included for completeness as an imaginary shift vector may be regarded
as unphysical. Nonetheless, noting that $\barbeta^2= \beta^2= -1$ on
$\Sigma$, the results \eqref{result3} and \eqref{wresult3} can be
connected by an algebraic maneuver on the shift vector. If we define
$\hat{\beta} \equiv \pm i\barbeta$ and $\tilde{\beta}= \pm i \beta$,
these changes could be seen as \textit{Wick rotations}, yielding
${\hat{\beta}}^2={\tilde{\beta}}^2=1$ on $\Sigma$.

For the inviscid general Burgers equation \eqref{generalburgers} to be
quasilinear hyperbolic it is necessary that
\be
c(\beta) = \beta^3 - \beta > 0,
\ee
which implies that $|\beta| > 1$ as a necessary condition for the 
particle inside the warp bubble to present superluminal wave-like behavior,
although, as mentioned above, this is just a possibility for superluminal
particle behavior.

An interesting additional result can also be obtained regarding heat and
thermal diffusivity. To reach this let us write Eq.\ \eqref{wresult1} as
follows,
\be
2 \frac{\partial\beta}{\partial t} - \nu \frac{\partial^2\beta}{\partial x^2} + \left[\nu \frac{\partial^2\beta}{\partial x^2} - \frac{\partial\beta}{\partial t} - \beta \frac{\partial \beta}{\partial x} \right] = - \beta^3 \frac{\partial \beta}{\partial x}, \label{newwresult1}
\ee
where $\nu$ is a real constant. It is possible now to define the following
two equations,
\begin{align}
F_1(t,\beta,\partial\beta/\partial x) &= \frac{\partial\beta}{\partial t} 
- \frac{\nu}{2} \frac{\partial^2\beta}{\partial x^2} \label{heateq}, 
\\
F_2(t,\beta,\partial\beta/\partial x) &= \frac{\partial\beta}{\partial t} 
+ \beta \frac{\partial \beta}{\partial x} - \nu \frac{\partial^2\beta}
{\partial x^2}, \label{burgerseq}
\end{align}
where $F_1$ is the heat equation with thermal diffusivity constant
$\nu/2$ and $F_2$ is the viscous Burgers equation with diffusion constant
$\nu$. Eq.\ \eqref{newwresult1} can then be rewritten as below,
\be
2 F_1 - F_2 = - \beta^3 \frac{\partial \beta}{\partial x}.
\ee
If both Eqs.\ \eqref{heateq} and \eqref{burgerseq} vanish, then the
solution of the heat equation $F_1$ and the viscous Burgers equation
$F_2$ mean  $\beta = 0$ or $\partial\beta / \partial x = 0$, which 
are two conditions consistent with the junctions conditions on $\Sigma$
matching the interior WD metric with the exterior Minkowski spacetime.  

\section{Flatness and vacuum conditions}\lb{flat}

Let us now discuss flatness and vacuum situations of the WD metric when
it reaches the boundary $\Sigma$.

\subsection{Shift vector $\barbeta$}

For the flatness case, considering the gauge given by Eqs.\
\eqref{result2} the nonzero components of the Riemann tensor on $\Sigma$
are reduced to the ones below,
\begin{subequations}
\begin{empheq}[left=\empheqlbrace]{align}
\mathrm{R}_{\phantom{t}t t x}^{t\phantom{t}\phantom{t}\phantom{x}} 
&=-\bar{\beta} \, \mathrm{R}_{\phantom{t}x t x}^{t\phantom{t}\phantom{t}
\phantom{x}}=-\bar{\beta} \, \frac{\partial}{\partial x}\left(
\frac{\partial \bar{\beta}}{\partial t}
+ \barbeta\, \frac{\partial\bar{\beta}}{\partial x}\right),
\label{rie1} \\[-8pt] \nonumber \\
\mathrm{R}_{\phantom{x}t t x}^{x\phantom{t}\phantom{t}\phantom{x}} 
&=-({\bar{\beta}}^2-1) \, \mathrm{R}_{\phantom{t}x t x}^{t\phantom{t}\phantom{t}
\phantom{x}}=-({\bar{\beta}}^2-1) \, \frac{\partial}{\partial x}\left(
\frac{\partial \bar{\beta}}{\partial t}
+ \barbeta\, \frac{\partial\bar{\beta}}{\partial x}\right).
\label{rie2}
\end{empheq}
\end{subequations}
Hence, the metric \eqref{ozwdmetric} becomes flat on $\Sigma$ either
when $\barbeta=0$, which is the uninteresting trivial case of the
junction condition \eqref{jc1} because then the WD spacetime is reduced
to the Minkowski one, or when the shift vector obeys the equation below,
\be
\frac{\partial \bar{\beta}}{\partial t}
+ \barbeta\, \frac{\partial\bar{\beta}}{\partial x}=0,
\;\;\; \mathrm{on} \;\;\; \Sigma.
\lb{burgers3}
\ee
Notice that the expression above is just the inviscid Burgers equation
\eqref{generalburgers} when $c(\barbeta)=\barbeta$. Further constraint
on $\barbeta$ is imposed at the joining hypersurface, because taking
together the definition \eqref{cbeta} and the inequality \eqref{inequa}
it is required that $\barbeta^3 \approx 0$ on $\Sigma$.

Therefore, the WD metric \eqref{ozwdmetric} is locally flat only once 
the inviscid Burgers equation is satisfied, that is, at the boundary 
hypersurface $\Sigma$ where the junction conditions are defined, beyond
which lies the exterior Minkowski spacetime.

To analyzed the vacuum case we require the components of the Ricci tensor.
Again, considering the gauge given by Eqs.\ \eqref{result2} the remaining
nonzero Ricci components on $\Sigma$ of the metric \eqref{ozwdmetric} yield,
\begin{subequations}
\begin{empheq}[left=\empheqlbrace]{align}
\mathrm{R}_{\:tt}&= \left(\bar{\beta}^2 - 1 \right)\mathrm{R}_{\:xx}
=\left(\barbeta^2-1 \right) \frac{\partial}{\partial x}
\left(\frac{\partial\bar{\beta}}{\partial t} + \barbeta \,
\frac{\partial \barbeta}{\partial x} \right),
\label{newricci00} \\[-8pt] \nonumber \\
\mathrm{R}_{\:tx} &=- \barbeta\, \mathrm{R}_{\:xx}
=-\barbeta \, \frac{\partial}{\partial x}
\left(\frac{\partial\bar{\beta}}{\partial t} + \barbeta \,
\frac{\partial \barbeta}{\partial x} \right).
\label{newricci01}
\end{empheq}
\end{subequations}
The vacuum is obtained with the trivial and uninteresting case of
$\barbeta=0$. However, the flatness condition given by Eq.\
\eqref{burgers3} also leads to vacuum. In addition, the special
result
\be
\barbeta=\pm1, \;\;\; \mathrm{on} \;\;\; \Sigma,
\lb{beta1}
\ee
not only implies on flat and vacuum results for the WD spacetime at the
joining surface, but also generates a \textit{singularity} on $\Sigma$
because the temporal part of the metric \eqref{ozwdmetric} vanishes. Refs.\
\cite[$\mathsection$ IV]{Alcubierre2017} and \cite[$\mathsection$ 2.3.2]
{teseosvaldo} provided the additional interpretation that the geometrical
pathology generated by the special result \eqref{beta1} may also mean the
formation of an event horizon in front and behind the warp bubble.

\subsection{Alcubierre shift vector $\beta$}

Considering the gauge given by Eqs.\ \eqref{wresult2}, the remaining
nonzero Riemann tensor components on $\Sigma$ of the metric
\eqref{alcmetric} are written below,
\begin{subequations}
\begin{empheq}[left=\empheqlbrace]{align}
\mathrm{R}_{\phantom{t}t t x}^{t\phantom{t}\phantom{t}\phantom{x}} 
&=-\mathrm{R}_{\phantom{t}x t x}^{x\phantom{t}\phantom{t}\phantom{x}} 
=\beta \mathrm{R}_{\phantom{t}x t x}^{t\phantom{t}\phantom{t}\phantom{x}} 
= -{\beta} \, \frac{\partial}{\partial x}\left(
\frac{\partial {\beta}}{\partial t} - \beta\, \frac{\partial{\beta}}
{\partial x}\right), \label{riealc1} \\[-8pt] \nonumber \\
\mathrm{R}_{\phantom{x}t t x}^{x\phantom{t}\phantom{t}\phantom{x}} 
&=-({\beta}^2-1) \, \mathrm{R}_{\phantom{t}x t x}^{t\phantom{t}\phantom{t}
\phantom{x}}=({\beta}^2-1) \, \frac{\partial}{\partial x}\left(
\frac{\partial {\beta}}{\partial t}
- \beta\, \frac{\partial{\beta}}{\partial x}\right).
\label{riealc2}
\end{empheq}
\end{subequations}
So, similarly to the previous case, flatness occurs in the trivial
situation of $\beta=0$, when
\be
\frac{\partial {\beta}}{\partial t}
- \beta\, \frac{\partial{\beta}}{\partial x}=0,
\;\;\; \mathrm{on} \;\;\; \Sigma,
\lb{burgers4}
\ee
and for $\beta=\pm1$, which also leads to a singularity in the metric
\eqref{alcmetric} at the boundary hypersurface $\Sigma$ possibly
interpreted as an event horizon. In this case $c(\beta)=-\beta$ for
the inviscid Burgers equation \eqref{generalburgers}.

One can also link the expressions above to the discussion of Sec.\
\ref{menosbeta} regarding heat and viscous components, as follows,
\be
\mathrm{R}_{\phantom{t}t t x}^{t\phantom{t}\phantom{t}\phantom{x}} 
= - \beta \, \frac{\partial}{\partial x}\left[2 F_1 - F_2\right],
\lb{f1f2}
\ee
where $F_1$ and $F_2$ are the heat equation and the viscous Burgers
equation, respectively defined in Eqs.\ \eqref{heateq} and \eqref{burgerseq}.

Concerning the vacuum case, the gauge given by Eqs.\ \eqref{wresult2}
produces the following nonzero components of the Ricci tensor on $\Sigma$,
\begin{subequations}
\begin{empheq}[left=\empheqlbrace]{align}
\mathrm{R}_{\:tt}&= \left({\beta}^2-1 \right)\mathrm{R}_{\:xx}
=\left(\beta^2-1 \right) \frac{\partial}{\partial x}
\left(\frac{\partial{\beta}}{\partial t} + \beta \,
\frac{\partial \beta}{\partial x} \right),
\label{newricci04} \\[-8pt] \nonumber \\
\mathrm{R}_{\:tx} &=- \beta\, \mathrm{R}_{\:xx}
=-\beta \, \frac{\partial}{\partial x}
\left(\frac{\partial{\beta}}{\partial t} + \beta \,
\frac{\partial \beta}{\partial x} \right).
\label{newricci05}
\end{empheq}
\end{subequations}
So, the original WD metric is then flat if $\beta = \pm 1$. Besides, at
the points of the spacetime where the heat equation and the viscous
Burgers equation are satisfied belong to the hypersurface that defines
the junction conditions with the exterior Minkowski spacetime.

\section{Conclusions} \lb{conc}

This paper analyzed the \textit{Darmois junction conditions} matching
the \textit{warp drive} (WD) metric interior spacetime with the exterior 
Minkowski geometry. The results show that a Burgers-equation-obeying
shift vector is needed to match continuously these two spacetimes on the 
joining hypersurface $\Sigma$, this being the case for both shift vectors 
studied here: the Alcubierre $\beta$ and its symmetric counterpart
$\barbeta$.

The \textit{first fundamental form} required for the match produces the
trivial and uninteresting cases of $\beta=\barbeta=0$. The \textit{second
fundamental form} yielded the results $\beta=\barbeta=\pm 1$ and
$\beta=\barbeta=\pm i$, which render the WD metric flat since $\beta$ and
$\bar{\beta}$ are constant and equal to the speed of light for the former
case, and regarded as unphysical for the latter one. The \textit{gauge}
$\partial\beta/\partial y=\partial\beta/\partial z = \partial\barbeta/
\partial y=\partial\barbeta/\partial z =0$ is also a required junction
condition produced by the second fundamental form, making vacuum solutions
possible for the WD and connecting them to the inviscid Burgers type
shock-waves.

A general inviscid Burgers equation of the type $\partial \barbeta/\partial
t + c(\bar{\beta})\, \partial \bar{\beta}/\partial x = 0$ was found as a
junction condition originated from the second fundamental form, having
$c(\bar{\beta}) = \bar{\beta} - \bar{\beta}^3$, which means that a
particle inside the warp bubble presents wave-like behavior for $0 <
|\bar{\beta}| < 1$. For the original Alcubierre shift vector $\beta$ we
found similar results in the form of a general inviscid Burgers equation
of the type $\partial \beta/\partial t + c(\beta) \partial \beta/\partial
x = 0$ as junction conditions, where $c(\beta) = \beta^3 - \beta$. The
particle inside the warp bubble presents wave-like behavior for $|\beta|
> 1$, so for the original $\beta$ hyper speed can be achieved with wave and
particle behavior. An important distinction between $\bar{\beta}$ and
$\beta$ is suggested that instead of a usual inviscid Burgers equation a
solution appears that is a linear combination of the heat equation with
$\nu/2$ as thermal diffusivity constant and the viscous Burgers equation
with diffusion constant $\nu$.

It was also shown that the Ricci and Riemann tensors are not constants
or zero on $\Sigma$ when the gauge is made for both $\beta$ and
$\bar{\beta}$. This means that \textit{there is surface gravity on the
interior side of the joining hypersurface.} The flatness of the WD
spacetime is only obtained at points where the inviscid Burgers equation
is satisfied for the WD metric with the shift vector $\bar{\beta}$, and
where the heat equation and the viscous Burgers equation are satisfied
for the original Alcubierre metric with $\beta$. Such points belong to
the junction conditions on the hypersurface $\Sigma$ that connects the
interior WD to the exterior Minkowski spacetime for both cases $\beta$
and $\bar{\beta}$. Hence, on the interior side of $\Sigma$ there still
is surface gravity, but on its outside the geometry is flat with no
gravity.

As final comments, similarly to what happens with the cosmological studies
mentioned at the Introduction, the match between an interior solution and
an external one is such that the external geometry is not altered by the
interior one. Therefore, for an outside observer ``sitting'' on the Minkowski 
spacetime it is as if the interior WD geometry were never there. This is so 
because in the interior WD spacetime the shift vector can be whatever the 
matter-energy distribution requires for generating warp speeds, but when the
shift vector reaches the matching boundary it must obey the inviscid Burgers 
equation so that both the Riemann and Ricci tensor components of the WD
metric vanish in order to avoid disturbing the exterior Minkowski geometry.
Then, as far as a possible superluminal speed navigation is concerned, the
task would be to produce an interior shift vector that not only obeys the
Burgers equation at the limits of the warp bubble, but also avoids creating
singularities at the matching wall in front and behind the warp bubble's
displacement.

\section*{Acknowledgments}
We are very grateful to Malcolm A.\ H.\ MacCallum for valuable comments
which pointed us in a direction that eventually resulted in this paper.
We also benefited from discussions with Miguel Alcubierre, Fernando Lobo
and Matt Visser. Thanks also go to a referee for useful comments. M.B.R.\
acknowledges partial financial support from FAPERJ -- \textit{Carlos 
Chagas Filho Foundation for the Support of Research  in the State of Rio de 
Janeiro} -- grant number E-26/210.552/2024.












\bibliography{warp}
\bibliographystyle{elsarticle-num}
\end{document}